\documentclass[11pt, conference]{IEEEtran}
\IEEEoverridecommandlockouts
\usepackage{cite}
\usepackage{amsmath,amssymb,amsfonts}
\usepackage{algorithmic}
\usepackage{graphicx}
\usepackage{textcomp}
\usepackage{xcolor}
\usepackage{balance} \usepackage{comment}
\usepackage{enumitem}

\def\BibTeX{{\rm B\kern-.05em{\sc i\kern-.025em b}\kern-.08em
    T\kern-.1667em\lower.7ex\hbox{E}\kern-.125emX}}
\begin{document}

\title{Optimization Framework for Green Networking}

\author{\IEEEauthorblockN{Cedric Westphal, Alexander Clemm}
\IEEEauthorblockA{
Futurewei Technologies,\\ Santa Clara, USA\\
\{cedric.westphal,alex\}@futurewei.com}
}

\maketitle

\begin{abstract}
Reducing energy consumption - and especially carbon emissions - is one of the most important challenges facing humankind. ICT (Information and Communication Technology) is a powerful tool to reduce emissions as it offers alternatives to activities that are costly in energy: video streaming saves energy vs driving to a movie theater, for instance. Still, the carbon footprint of ICT in general and networking in particular have been growing, despite better energy efficiency per transmitted bit, due to the sheer growth in Internet usage and traffic. 

The information and communication technology (ICT) sector is currently estimated to create 2.7\% of all global CO2 emissions ~\cite{lorincz2019greener} and expected to continue to increase. Hence, monitoring and reducing the CO2 emissions from ICT is increasingly important. Networks are responsible for
around 13\% of ICT energy consumption~\cite{andrae2020new}, a third of which in turn is attributable to backbone (core) networks~\cite{hinton2011power}. As such, it is important to offer new mechanisms to reduce the energy footprint of networks. 

In this paper, we present a framework to include energy considerations (as a proxy for carbon footprint) into the management plane of a network. We apply this framework to optimize the network topology so as to minimize the energy spending while at the same time providing a satisfactory Quality of Experience (QoE) to the end users.  We present this framework along high-level considerations, as its deployment and evaluation is left for future work. 
\end{abstract}


\section{Introduction}
\label{sec:intro}

Climate change and the need to curb greenhouse emissions have been recognized by the United Nations and by most governments as one of the big challenges of our time. As a result, improving energy efficiency and reducing power consumption are becoming of increasing importance for society and for many industries. The networking industry is no exception.

Information and Communication Technology (ICT) in general can be viewed as a tool to save energy. It is for instance understood that a virtual conference consumes a fraction of the energy cost of an in-person conference, where the main budget is air travel. For the IETF standardization organization meetings, which tabulated this in an effort to reach Net Zero emission~\cite{IETFNetZero}, the cost of air travel amounts to 99\% of the total energy cost of an event.

While ICT saves energy by such substitution effects, networks themselves still spend a lot of energy and a lot of effort goes into making them more efficient. Telefonica~\cite{telefonica2021} reports that in 2021, its network's energy consumption per PB of data added up to 54 MWh.  
This amount has dramatically decreased by a five-fold factor over the previous five years. However, gains in efficiency are quickly offset by simultaneous growth in data volume. This report states a goal to reduce carbon emissions by 70\% over the next five years.

There are two ways to reduce carbon emissions: switching to clean energy sources, which is outside of the scope of this paper; and reducing the consumption of energy of the network, which we focus on here. 

In typical networking devices, only roughly half of the energy consumption is associated with the data plane~\cite{bolla2011energy}. Turning the system on consumes more than half of the power before any traffic is transmitted when compared with same system operating at full load~\cite{chabarek08,cervero19}. A device's power consumption does not grow linearly with the volume of forwarded traffic. It looks more like a step function: the cost of the first bit is very high, as it requires powering up a device, port, etc. The marginal cost of transmitting more traffic is then very low.  Likewise, the energy cost of incremental CPU and memory needed to process additional packets becomes negligible.

In other words: networking equipment today is not energy proportional. The amount of energy spent to transmit data over a link is not proportional to the amount of data being transmitted. While energy-proportionality should be a focus of the systems' research community to achieve greater energy efficiency, until such networking hardware is deployed, we need to focus on how we can intelligently turn resources (in general) and links (in particular) on and off. By turning resources off when they are not needed, we can expect to affect energy spending in meaningful ways. Of course, we need to do so in ways that do not adversely affect users' Quality of Experience (QoE) and other important network properties such as resilience. 

We present here a framework to optimize the energy consumption of the network. We consider a system where network elements create a data plane to forward the data, and are connected to a control plane for network management.  We leverage the fact that energy consumption is not energy proportional, hence energy can be saved by ensuring high link utilization (e.g. batching communication up where possible) and switching links off in the periods in between.

In our framework, the controller collects information about energy consumption and transmitted data from the forwarding elements, use this as input to an optimization problem. The results of the optimization comprise turning off as many links as possible to configure an active network topology that is as green as possible, while at the same time satisfying some minimal level of QoE for the users.

One key element of our framework is the observation that a large amount of Internet traffic is composed of rate adaptive video traffic~\cite{ciscoForecast}: this means that the applications of the network can respond to routing by adapting the session rate and the total amount of data being transmitted. 

For instance, if we have two nodes A and B and two links $l_1$ and $l_2$ of capacity one linking A and B, and the users of the network request two  DASH~\cite{DASH} video streams $s_1$ and $s_2$ between the clients attached to A and the video server at B. Further, the DASH sessions can take rate of $\{0.25,0.5,1\}$ units per second.

If both links are up, then $s_1$ can be assigned to $l_1$, $s_2$ to $l_2$ and each link can transmit at the highest rate of 1. The total amount of data transmitted is 2 units per second.

\begin{figure}
    \centering
    \includegraphics[width=0.5\textwidth]{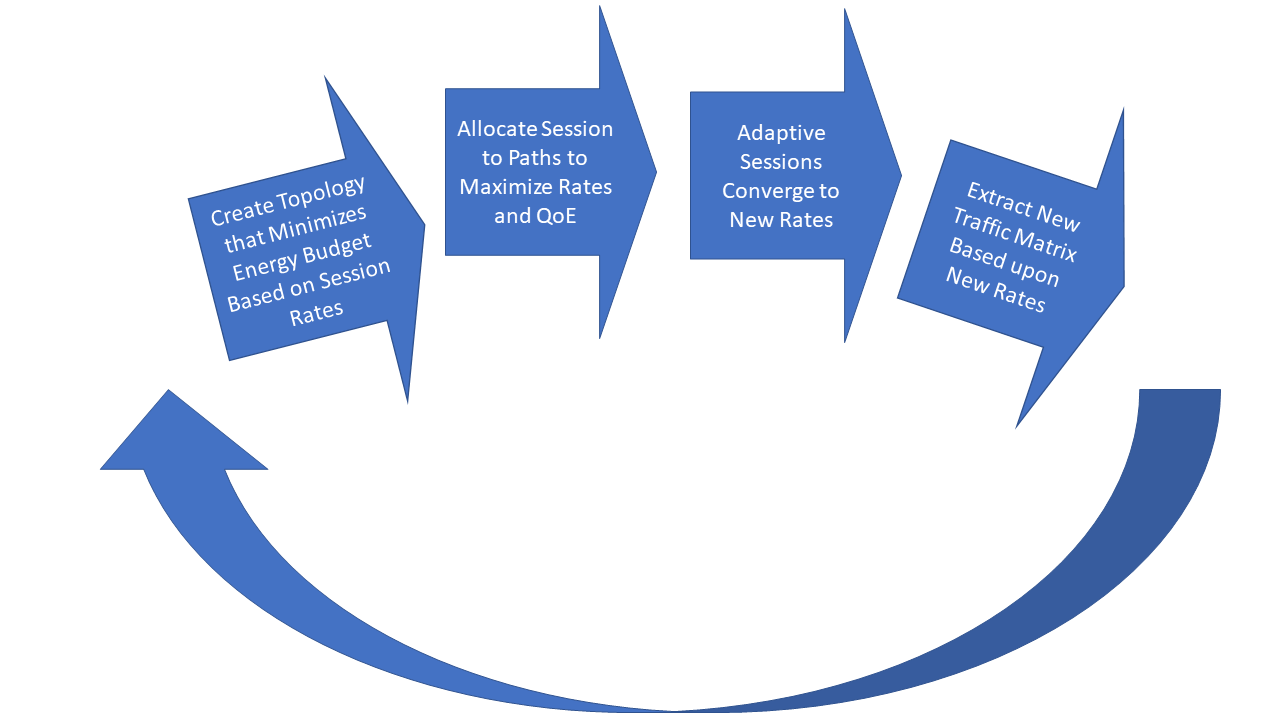}
    \caption{Joint Energy and QoE Optimization}
\label{fig:loop}
\end{figure}

The network can also decide to turn $l_2$ off. Now both $s_1$ and $s_2$ use $l_1$ and their rate converges to $0.5$, as the capacity of $l_1$ is 1. The total amount of data transmitted is half of the previous scenario, that is 1 unit per second. The energy consumption is also half in this case. However, there is a performance degradation from the point of view of the users, who receive poorer QoE from the lower rate. If the capacity of $l_1$ and $l_2$ is 2 units per link, then one can be turned off without any performance degradation: both sessions get 1 unit per unit of time. 

It makes therefore sense to turn links off to reduce energy consumption, as long as the gain from the reduction in energy is not offset by such a loss of QoE that the end users would not tolerate. The task is therefore to figure out the trade-off between turning off part of the infrastructure while still delivering enough to satisfy the network users. We present a framework that attempts to solve this trade-off in a satisfactory manner. This is illustrated in Fig.~\ref{fig:loop}.

This paper is organized as follows: Section~\ref{sec:rel} presents some background on network energy efficiency. Section~\ref{sec:problem} describes our problem formulation. Section~\ref{sec:algo} describes approaches to solve this problem. Section~\ref{sec:system} show how to deploy this into a practical network with distinct control and forwarding planes. Section~\ref{sec:next} describes future steps for this work and offers some concluding thoughts.


\section{Related Work}
\label{sec:rel}

Green networking becomes more and more critical. Some work on this topic has been proposed into IETF (see, for instance,~\cite{cx-green-ps-01}).  Some of the challenges brought on by trying to green the network are described in \cite{clemm2022challenges}, along with a list of potential research directions. This is a good overall overview of green networking. 

Previous documents~\cite{bolla2011energy,lorincz2019greener,bianzino2012survey,samdanis2015green} survey the trends for a greener Internet in the Internet and some of the proposed solutions. \cite{davaslioglu20165g} is another survey focusing on the impact of 5G for green networking. 

A different trade-off is considered in \cite{welzl2022tradeoff}: it looks at links going on and out of a low-energy sleep mode by transmitting data in batch. This is applied to WiFi links and shows that there is a potential energy gain in transmitting data more aggressively to complete faster and return to a low power mode. Our approach is a network-wide approach that leverages the elasticity of rate-adaptive streaming, but both are inspired by the same philosophy. 

Some of the challenges overlap with that of~\cite{westphal2017challenges}: network support for AR/VR would offer a competitive substitute solution to a lot of in-person meetings, and it is known that air travel is the main energy cost of a business meeting, as mentioned above.

Another proposal leverages the SCION protocol to make network paths aware of carbon emissions \cite{tabaeiaghdaei2022carbon}. For instance, routing could select paths with low carbon emissions due to the use of renewable energy. This is complementary with our approach.

The approach of this paper is similar to~\cite{ren2018jordan}. However, this work did not consider energy as a cost, which leads to significant differences that this paper addresses. In particular, it considers as cost as a function of the amount of traffic going through a link. Here, the energy is a step function: either a link is up, and consuming energy independently of its utilization/traffic; or it is down, and its consumption is 0.  

\section{Problem Formulation}
\label{sec:problem}

\subsection{Network Model}
\label{sec:net}

We consider a network graph $G(V,E)$ where the vertices $V$ are connected by edges in $E$. $V$ is the set of nodes in the network, and $e_{i,j}$ is the link from node $i$ to node $j$. Each link has a capacity $c_{i,j}$. The link utilization $l_{i,j}$ is equal to the bandwidth allocated to link $e_{i,j}$ divided by the capacity $c_{i,j}$. 

We consider an energy model where each link is on, and consumes an amount of energy $\epsilon_{i,j}$. This amount is zero if the link if off. We assume the link is off if it carries no traffic. 

Customers (or users) want to exchange data over the network. To do so, they initiate session $k$ with bitrate $r_k$, where $k$ is the highest achievable vector drown from $Q$, where $Q$ is the set of possible rate vectors for the sessions. If the session is not elastic, then $Q$ contains only one vector. If the session is for adaptive video streaming (as most of the traffic on the Internet currently is), then $Q$ contains a discrete list of possible rates. 

$k$ has a source and a destination within the network graph $s$ and $d$ in $V$ respectively. Under a set of requests from the users, the network needs to allocate the traffic to the different possible paths between each (source,destination) pairs for each user $k$. $r_{i,j,k}$ is the amount of traffic on link $e_{i,j}$ for session $k$.

We consider the users derive a utility $U(r_k)$ for session $k$, where $U$ is the user benefit function that is a positive, concave, positive function. The sum of $U(r_k)$ for all users $k$ is the total user benefit. 

We define the energy cost of an allocation. Denote by $\mathbb{1}$  the indicator function that is equal to 1 if its input is true, and 0 otherwise. Then the energy cost of a link $e_{i,j}$ is $f_{i,j} = \epsilon_{i,j} \mathbb{1}\{\Sigma_k r_{i,j,k}>0\}$. We can also equivalently write $f_{i,j} =  \epsilon_{i,j} \mathbb{1}\{l_{i,j}>0\}$.

We would like to minimize the energy cost while delivering a satisfying experience to the users. This means that we would like to jointly optimize for maximizing $\Sigma_k U(r_k)$ while minimizing $\Sigma_{i,j} f_{i,j}$.

Two possible approaches are therefore:
\begin{itemize} 
\item jointly minimize
\begin{equation}
\label{eq:min1}
     -\alpha \Sigma_k U(r_k) + \beta \Sigma_{i,j} f_{i,j}
\end{equation}
for two positive coefficients $\alpha$ and $\beta$ chosen to properly weight the two objectives. 
\item Set a value for the minimal desired QoE of the users, namely $U(r_k) > U_w$ and then minimize:
\begin{equation}
\label{eq:min2}
    \min \Sigma_{j,k} f_{i,j} \mbox{ under the constraint: } U(r_k) > U_w
\end{equation}
\end{itemize}

The first one may find a higher utility at the cost of sacrificing some  (while providing satisfying QoE {\em on average}). The second one sets a bottom QoE that all users need to achieve.

Note that we can also set the rate vector to only include acceptable rates. For instance, if we need the rate to be greater than $r_min$, then we include in $Q$ only rates greater than $r_min$. This way, solving Problem 1 will yield a minimal QoE level for every user. 

\subsection{Constraints}
\label{sec:constraints}

Recall that $r_{i,j,k}$ is the rate of session $k$ going through link $e_{i,j}$. The link capacity constraint is therefore:
\begin{equation}
    \Sigma_k r_{i,j,k} \leq c_{i,j} \forall e_{i,j} \in E \label{eq:cap}
\end{equation}

The flow conservation constraint is:
\begin{eqnarray}
\Sigma_{e_{i,j} \in \Gamma^+(v)} r_{i,j,k} - \Sigma_{e_{i,j} \in \Gamma^-(v)} r_{i,j,k} = 0, \\ \nonumber
\forall k \in K, v \neq s_k,d_k \label{eq:flow1} \\
\Sigma_{e_{i,j} \in \Gamma^+(v)} r_{i,j,k} = r_k, \\ \nonumber
\forall k \in K, v=s_k \label{eq:flow2} \\
\Sigma_{e_{i,j} \in \Gamma^-(v)} r_{i,j,k} = r_k, \\ \nonumber
\forall k \in K, v=d_k \label{eq:flow3}
\end{eqnarray}
where $\Gamma^+(v)$ is the set of incoming links to $v$ and $\Gamma^-(v)$ is the set of outgoing links.

The problem becomes: 

\textbf{Problem 1:} Minimize (\ref{eq:min1}) under Link Capacity Constraint~(\ref{eq:cap}), Flow Conservation Constraints~(\ref{eq:flow1})-(\ref{eq:flow3}) and Variables $r_k \in Q, r_{i,j,k} \geq 0$.

\section{Algorithm Design}
\label{sec:algo}

Solving Problem 1 is difficult, as the rate vector $Q$ is not linear, and as the energy cost is not concave. 

We therefore propose a heuristic solution to a different problem, where we relax some of the requirements of Problem 1.

The flow conservation constraint at the destination is redundant, and we can replace Equation~\ref{eq:flow2} with an inequality:
\begin{eqnarray}
\Sigma_{e_{i,j} \in \Gamma^+(v)} r_{i,j,k} \geq r_k, \forall k \in K, v=s_k \label{eq:flow4}
\end{eqnarray}

We can move Equation~\ref{eq:flow4} into the objective:
\begin{eqnarray}
L(\lambda) & = & -\alpha \Sigma_k U(r_k) + \beta \Sigma_{i,j} f_{i,j} + \\ \nonumber
& & \Sigma_k \lambda_k (r_k - \Sigma_{e_{i,j}} r_{i,j,k} r_{i,j,k} ) \label{eq:flow5}\\
& = & \beta \Sigma_{i,j} f_{i,j} - \Sigma_k \Sigma_{e_{i,j}} \lambda_k r_{i,j,k} \\ \nonumber
& & -\alpha \Sigma_k U(r_k) + \Sigma_k \lambda_k r_k \label{eq:flow6}
\end{eqnarray}

Since the first line of Equation~\ref{eq:flow6} depends only on the $f_{i,j}$ and $r_{i,j,k}$ and the second line only on $r_k$ we can separate our initial problem into two sub-problems:

\begin{itemize}
    \item {\bf Sub-problem 1:} Minimize $\beta \Sigma_{i,j} f_{i,j} - \Sigma_k \Sigma_{e_{i,j}} \lambda_k r_{i,j,k}$ under constraints under Link capacity constraint~\ref{eq:cap}, flow conservation constraint~\ref{eq:flow1} and $r_{i,j,k} geq 0$
    \item {\bf Sub-problem 2:} Minimize $-\alpha \Sigma_k U(r_k) + \Sigma_k \lambda_k r_k$ under constraints $r \in Q$.
\end{itemize}

We can linearize $r_k$ and then select the nearest lower rate within the vector $Q$.

This is a well defined problem for which an approximation can be found using classical methods. Therefore, if the inputs to the problem are provided to us, we are able to find out a solution. 

\section{System Design}
\label{sec:system}

We need to describe the overall system that will leverage the method of the previous Section. 

Our framework centers around a controller that has a complete view of the network, including the links and their energy characteristics, as well as an indication of the traffic demand.  Using those inputs, the controller calculates the optimal topology at any one point in time, solving the optimization problem of minimizing energy use while satisfying the constraints on QoE that still needs to be delivered. 

Specifically, the calculation involves determining which links can be turned off selectively, which link traffic rates should be adjusted, and how routing / paths should be configured. Subsequently, the controller provisions the network (links and routing tables) accord

We consider the following system comprised of:
\begin{itemize}
    \item A network of forwarding elements;
    \item Connected by a set of links with known capacity and known power consumption;
    \item That interface with a logically centralized controller - as in SDN or in any other control plane method;
    \item Using a reporting mechanism (a control protocol for instance, either in-band or out-of-band) where the network elements report their link status, link utilization and/or link power consumption to the controller;
    \item A control protocol where the controller can turn on/off;
    \item A routing protocol where the controller populates the route tables in the forwarding elements;
    \item A function in the controller that takes as input the traffic into the network (as a set of rate-adaptive sessions) and calculates the rates to be assigned to the sessions (when elastic or adaptive), the total resulting demand, and the required links to be turned on or off to support this demand as an output of the algorithm of the previous Section.
\end{itemize}

In other words, a controller takes as input the traffic entering the network over a period of time (described as a set of sessions with potentially an associated set of rates). Then based upon the energy cost of turning on/off links in the network, the controller is able to calculate a set of active links to turn on, a set of stand-by links to put into sleep mode (or turn off), and a routing table. This is achieved using the method of Section~\ref{sec:algo}. The controller is then able to push this routing table at the forwarding elements and to gather the statistics from these elements for the next period of time. 

\begin{figure}
    \centering
    \includegraphics[width=0.5\textwidth]{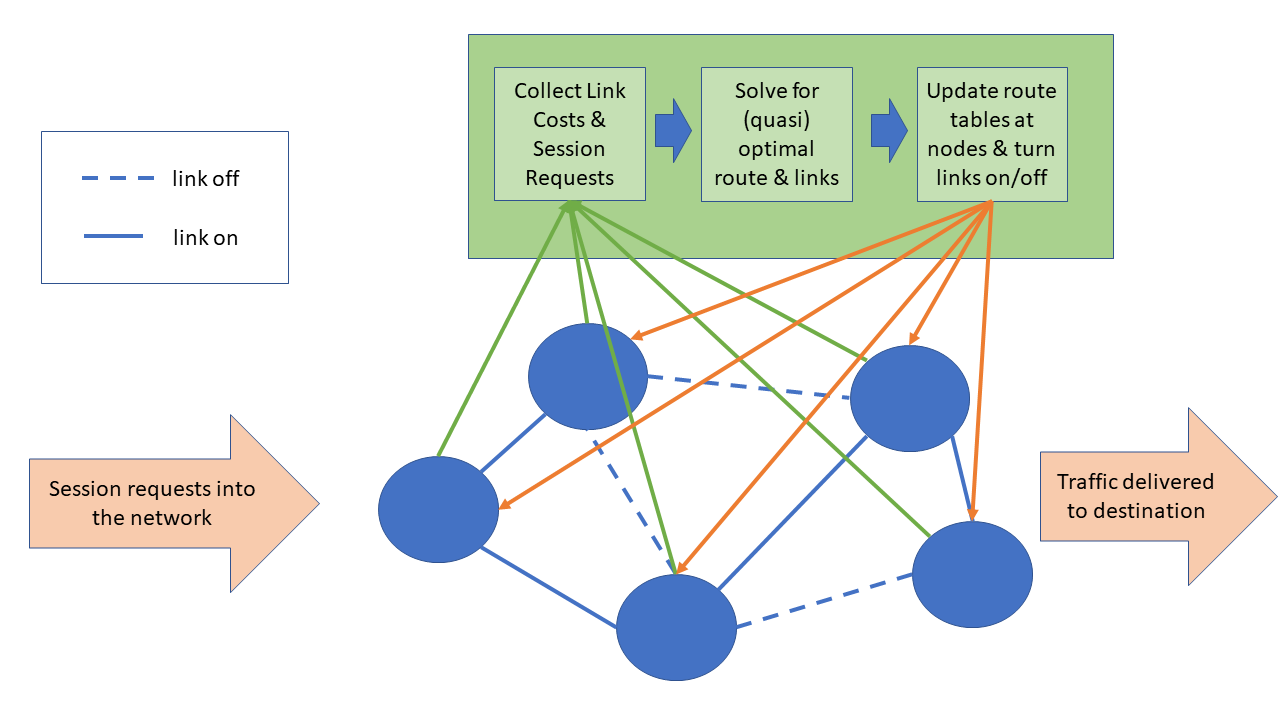}
    \caption{System Overview}
    \label{fig:system_overview}
\end{figure}

Figure~\ref{fig:system_overview} presents an overview of the system design.

\section{Next Steps \& Conclusions}
\label{sec:next}

This paper presents a framework from 30,000 feet, without getting into the implementation details or the evaluation of the performance.

Critical to the next steps is to materialize this framework into a simulated and/or implemented systems. This would allow us to benchmark the system and to assess its potential in turning links on and off and in quantifying the savings in energy.

For future work, we will evaluate the proposed optimization model on a series of representative topologies. 

One critical factor is the support from the infrastructure for such proposal: namely the ability to turn links (or even whole network cards) on and off quickly, and to save energy doing so. We need also infrastructure support to monitor and report the energy consumption. As emphasized in \cite{clemm2022challenges}, one foundational building block concerns better instrumentation of the infrastructure to include visibility into energy-related matters into the management plane. Our proposal is aligned with this need, as it leverages such instrumentation and such energy knobs in the control plane.

We are also interested in deploying these ideas into actual networks. SDN provides the basic abstraction to support our framework: forwarding elements, controller, northbound interface to report on traffic demands and energy costs, protocols to set routing table and potentially put link into sleep mode. We are eager to try our proposals in such an environment.

Another direction for future steps is to push towards standardization of such proposal. An Internet Draft currently under discussion \cite{cx-green-ps-01} points out to the challenges that a standardization body such as IETF should consider. The IAB-organized workshop Environmental Impact of Internet Applications and Systems\cite{IABworkshop} points to an interest towards defining and standardizing practical solutions into a suite of inter-operable protocols. 

\balance

\bibliographystyle{IEEEtran}
\bibliography{GreenNet}

\end{document}